# Tuning quantum channels to maximize polarization entanglement for telecom photon pairs

DANIEL E. JONES,* BRIAN T. KIRBY, AND MICHAEL BRODSKY

U.S. Army Research Laboratory, Adelphi, MD 20783, USA
*Corresponding author: daniel.e.jones161.civ@mail.mil

**Quantum networks entangle remote nodes by distributing quantum states, which inevitably suffer from decoherence while traversing quantum channels. Pertinent decoherence mechanisms govern the channel capacity, its reach, and the quality and rate of distributed entanglement. Hence recognizing, understanding, and modeling those mechanisms is a crucial step in building quantum networks. Here, we study practical fiber-optic quantum channels that partially filter individual modes of transmitted polarization entangled states and are capable of introducing dephasing. First, we theoretically model and experimentally demonstrate the combined effect of two independent and arbitrarily oriented polarization dependent loss elements experienced by each photon of an entangled photon pair. Then, we showcase the compensation of lost entanglement by properly adjusting the channels' properties and discuss the resulting tradeoff between the entanglement quality and rate. Our results provide insights into the capacity of practical fiber-optics channels, thus taking an important step towards the realization of quantum networks.**

## 1. INTRODUCTION

Rising interest in quantum networks is fueled by the well-recognized prospects of powerful algorithms and protocols for secret sharing, distributed computing, authentication, and security. These promising functionalities rely on establishing high fidelity quantum entanglement between distant nodes of a quantum network [1]. Recent entanglement distribution field trials over deployed fiber [2,3] and free space channels [4] have just begun to probe the realm of rates and distances that are feasible with current technology. At the same time, novel approaches to improve the reach are being investigated in long distance lab testbeds and include ultra-bright sources [5], waveguide-based unbalanced interferometers [6], highly efficient superconducting detectors [6-10], and multi-dimensional entanglement together with high visibility Franson interferometry [9]. Furthermore, newly proposed protocols such as heralded steering [8] and two-hierarchy entanglement swapping [11] lead to higher tolerance of loss and noise.

On the other hand, the fundamentals of quantum channel capacities are being explored theoretically for a number of canonical quantum channel models such as pure loss bosonic channels [12, 13] and depolarizing [14], dephasing, and erasure channels [12, 14]. Active topics include channel robustness and error sources [15, 16], methods for estimating channel capacity [17], and tradeoffs between loss and rate [18] or rate and fidelity [14]. All of these exemplary channels could be effectively simulated with linear optics; for instance, to demonstrate delayed choice decoherence suppression [19], to simulate interactions with a boson bath [20], to measure the capacity of a channel [21], or to probe channel decoherence effects on a qubit in a cluster state [22].

An interesting open question is the capacity of practical channels, which may not necessarily always be described by the models above. Answering this question requires understanding of the pertinent decoherence mechanisms as well as their effect on the rate and quality of distributed entanglement. For instance, polarization entangled photons propagating in optical fibers are susceptible to polarization mode dispersion (PMD) [23-25] and polarization dependent loss (PDL).

Quantum mechanically, PDL could be viewed as a local filtering operation [26], which received some attention in the past in the context of entanglement distillation [27-29]. Experimentally, properly applied procrustean local filtering has been shown to increase the quantum correlations of certain classes of mixed states [30, 31] in free space table top experiments. The fiber-optic environment differs from free space optics in a significant fashion. Due to small intrinsic residual birefringence, the orientation of PDL vectors in fiber-optic channels could be arbitrary and dependent on the changing ambient conditions. Hence, there is a need for a concise and comprehensive general description of the arbitrarily oriented PDL, which is the main goal of our paper.

In this article, we devise a concise and elegant model describing how the concurrence of a general two qubit Bell Diagonal (BD) state depends on arbitrarily oriented PDL elements acting on each qubit. For an entanglement distribution system comprised of a source and two quantum channels, our model establishes the equivalence mapping between PDL elements in different channels. To verify the theory, we experimentally study the PDL resilience of the two specific BD states that are most relevant to the fiber-optic environment – a nearly perfect $|\Phi^+\rangle$ Bell state and a rank two Bell Diagonal state. The latter results from a $|\Phi^+\rangle$ Bell state partially decohered by PMD. Our theory and data show that when PDL exists in one channel only, the entanglement of the pair is affected by the magnitude but not the Stokes orientation of the PDL element. More generally, when an arbitrarily aligned PDL element is present in each channel, we find that the average entanglement only depends on the arithmetic sum of two PDL magnitudes. Our equivalence mapping permits an easy way to consider the relative orientation of the PDL elements, each belonging to a different channel. Although the relative orientation does not change the average entanglement, it does result in a tradeoff between the entanglement quality and the transmission rate. Therefore, the introduction of a single PDL element in only one channel can compensate for PDL present in either or both channels. We demonstrate this experimentally for a $|\Phi^+\rangle$

Bell state degraded by PDL in one channel with and without additional PMD impairment.

## 2. EXPERIMENTAL SETUP

The schematics of our experiment are illustrated in Fig. 1(a). Our fiber-optic testbed consists of an entangled photon source (EPS) and two separate detector stations (DS) [32]. Additionally, PDL emulating/compensating modules (PDLE/PDLC) can be inserted into the photon paths of channels A and B as needed. All of our modules are fully tunable in both PDL magnitude (0dB – 7dB) and Stokes orientation. One of the emulators (PDLE w/PMD) has an additional differential group delay of τ=6.6ps.

Inside the EPS, a 50MHz pulsed fiber laser operating at 193.1THz creates signal and idler photons via four-wave mixing by pumping a dispersion shifted fiber (DSF) [33]. The DSF is arranged in the Sagnac configuration using a polarization beam splitter (PBS). Depending on the direction in which the pump pulse creates the signal and idler photons, they can both be either vertically $|VV\rangle$ or horizontally $|HH\rangle$ polarized. Setting the pump polarization to 45° removes the directional information and results in a Bell state encoded in polarization $|\Phi^+\rangle = \frac{1}{\sqrt{2}}(|HH\rangle + |VV\rangle)$. At the output of the Sagnac loop, a WDM demux filters out the pump and separates photons spectrally into 100GHz-spaced ITU outputs [34]. For this experiment, we use ITU channel 28 (192.8THz) for quantum channel A and ITU channel 34 (193.4THz) for quantum channel B. The average number of generated photon pairs per pulse µ is tunable in the 0.001 – 0.1 range [35, 36]. Upon passing through quantum channels A and B, each photon reaches a detector station consisting of a polarization analyzer (PA) and an InGaAs single photon detector (SPD). The detectors are biased to operate with a detection efficiency of η ∼ 20% and a nearly negligible dark count probability of about 4x10$^{-5}$ per gate. Fully automated FPGA-based controller software performs full polarization state tomography [37, 38] and two-photon interference measurements.

Our source generates a nearly perfect two photon $|\Phi^+\rangle$ state with fidelity of about 0.95. To characterize it, we connect both detector stations directly back-to-back (B2B) to the source, perform state tomography measurements, and then calculate the concurrence of the resulting density matrix. The B2B concurrence varies slightly between different days and power off/on cycles. Overall, the B2B concurrence falls within the tight interval of $0.925 \pm 0.008$. The deviation from unity occurs mostly due to the presence of noise photons generated in the DSF by Raman scattering and pump leakage into the entangled photon band [34, 39]. Absolute values of a typical density matrix obtained by the B2B state tomography measurements are shown in Fig. 1(b). The plot reveals a slight imbalance between the $|HH\rangle$ and $|VV\rangle$ modes, which is constantly present. The ratio of the $|HH\rangle\langle HH|$ and $|VV\rangle\langle VV|$ matrix elements is about 1.38 (or 1.4 dB). We attribute this asymmetry to the specification tolerances of the off-the-shelf components comprising the EPS.

The state produced by the EPS could be viewed as an *ideal* Bell state $|\Phi^+\rangle = \frac{1}{\sqrt{2}}(|HH\rangle + |VV\rangle)$ that undergoes external mode damping by a *virtual source PDL* element whose axis of maximal transmission is aligned to the $|H\rangle$ polarization, or in other words, points in the $\sigma_3$ direction in Stokes space. For the special case of such alignment between the PDL element and the basis of the input Bell state, there is ambiguity in the exact placement of the source PDL element. Indeed, positioning the source PDL in either of the channels as indicated by $\vec{\gamma}^S$ in Fig. 2(a, b) would result in the experimentally observed imbalanced density

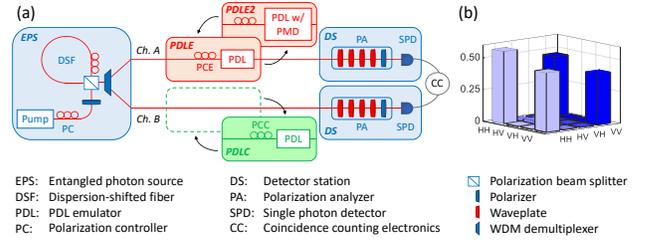

**Fig. 1** (a) Experiment schematics. (b) Absolute value of back-to-back density matrix with PDLE/PDLC removed.

matrix. In fact, any combination of PDL elements in channels A and B would produce the same state as long as the PDL elements are aligned to $|H\rangle$ and their PDL values add up to 1.4 dB. For the more general case of an arbitrarily oriented PDL element acting on a Bell state, *the equivalence mapping* of Fig. 2(a, b) becomes more involved.

## 3. DECREASE OF ENTANGLEMENT DUE TO POLARIZATION DEPENDENT LOSS

For our theoretical analysis, we consider the action of PDL operators $P_A$ and $P_B$ in channels A and B on the density matrix $\rho$ of the initial state [27-29,40]: $(P_A \otimes P_B)\rho(P_A \otimes P_B)^\dagger$. To simplify treatment of the PDL orientations, we choose the rotational form of the PDL operator [41]:

$$P = e^{-\frac{\gamma}{2}} e^{\frac{1}{2}\vec{\gamma}\cdot\vec{\sigma}} = e^{-\frac{\gamma}{2}}\left(\sigma_0 \cosh\left(\frac{\gamma}{2}\right) + (\hat{\gamma} \cdot \vec{\sigma}) \sinh\left(\frac{\gamma}{2}\right)\right), \quad (1)$$

where $\vec{\gamma}$ is the PDL Stokes vector, $\sigma_0$ is the 2x2 identity matrix, and $\vec{\sigma}$ is a vector of the Pauli matrices. Here $\vec{\gamma} = \gamma\hat{\gamma}$, where $\hat{\gamma}$ is the unit Stokes vector in the direction of maximum transmission and $\gamma$ is the PDL magnitude, which is related to its dB value by $\gamma = \frac{PDL_{dB}}{20 \log_{10} e}$. For the initial matrix $\rho$, we select a general Bell diagonal state which is of the form:

$$\rho = \left(\sigma_0 \otimes \sigma_0 + \sum_{j=1}^{3} t_j \sigma_j \otimes \sigma_j\right), \quad (2)$$

where $\sigma_0$ is the two dimensional identity matrix, $\sigma_j$ are again the Pauli matrices, and the diagonal matrix $T = diag(t_1, t_2, t_3)$ is referred to as the correlation matrix. Our primary focus is on Bell states, which all satisfy the constraint that $|t_j| = 1$. However, we choose the more general case of Bell diagonal states ($|t_j| \leq 1$) since the algebraic form of our results remains the same, and an analysis of the Bell diagonal states encapsulates the potential effects of noise and decoherence.

Utilizing the operator $P$ of Eq. 1, the initial state $\rho$ of Eq. 2, and the previous results of [28, 29] we derive the concurrence of the state affected by PDL $\rho' \propto (P_A \otimes P_B)\rho(P_A \otimes P_B)^\dagger$:

$$C(\rho') = \frac{C(\rho)}{\cosh(\gamma_A)\cosh(\gamma_B) + \kappa \sinh(\gamma_A)\sinh(\gamma_B)}, \quad (3)$$

where $\gamma_A$ and $\gamma_B$ are the magnitudes of the PDL in channels A and B respectively, $C(\rho)$ is the concurrence of the initial state, and $\rho'$ is the state after PDL is applied. Here we define the function

$$\kappa = \sum_{j=1}^{3} \gamma_{Aj}\gamma_{Bj}t_j = (T\hat{\gamma}_A) \cdot \hat{\gamma}_B = \hat{\gamma}_A \cdot (T\hat{\gamma}_B), \quad (4)$$

where $\gamma_{Aj}$ ($\gamma_{Bj}$) is the $j$th component of the PDL Stokes unit vector $\hat{\gamma}_A$ ($\hat{\gamma}_B$) of channel A (B). Note that $\kappa$ compactly describes the complex dependence of the concurrence of the final state on the relative orientation of the two PDL elements.

We start with an experimental setup that has only one PDL module inserted into channel A, shown as the red PDLE box in Fig. 1(a). We sequentially set the emulator to five different PDL magnitudes of 1.25, 2.55, 3.7, 5.1, and 6.3 dB. At each PDL value, the polarization controller PCE goes through 50 different arbitrary settings such that the resulting PDL Stokes vector $\vec{\gamma}^E$ covers the entire Poincaré sphere. Full tomography is performed at each setting. From each of the 250 experimentally obtained density matrices, we compute the state's concurrence and purity. There are two ways to present the concurrence data depending on the placement of the virtual source PDL $\vec{\gamma}^S$ depicted in Fig. 2(a, b), allowing us to verify Eq. 3 in two different scenarios.

When the virtual source PDL $\vec{\gamma}^S$ and the emulator PDL $\vec{\gamma}^E$ are in the same channel as depicted in Fig. 2(a), the PDL of channel B is zero. In this case, the concurrence of Eq. 3 reduces to a simple orientation-independent form given by:

$$C(\rho') = \frac{C(\rho)}{\cosh(\gamma_A)}, \quad (5)$$

where $\gamma_A$ is the magnitude of the aggregate PDL $\vec{\gamma}_A$ in channel A. The magnitude $\gamma_A$ can be extracted from a properly rotated experimental density matrix for each of our 250 tomography data points. Fig. 2(c) plots the experimental concurrence C vs the extracted values of $\gamma_A$ (shaded circles) along with the theoretical dependence of Eq. 5 (black line). The figure shows nearly perfect agreement between experiment and theory. In general, the orientation $\hat{\gamma}_A$ and the magnitude $\gamma_A$ of the aggregate PDL $\vec{\gamma}_A$ are governed by the cumbersome concatenation rules of the two PDL vectors $\vec{\gamma}^S$ and $\vec{\gamma}^E$ [42-46]. These rules allow us to compute the angle $\vartheta$ between vectors $\vec{\gamma}^S$ and $\vec{\gamma}^E$ from the measured value $\gamma_A$, which we use below. Experimentally, by controlling $\vartheta$ we sample a wide range of $\gamma_A$ values. This also ensures that the orientation of the aggregate PDL $\hat{\gamma}_A$ changes dramatically between the data points shown on the plot. This verifies that the concurrence is independent of the orientation $\hat{\gamma}_A$ of the aggregate PDL. To address the vertical spread in the data, we shade the symbols by their corresponding purity values with lighter colors corresponding to higher purity. The source performance varies slightly from day-to-day and between power cycles. Those variations result in the spread, since data points with a slightly higher purity naturally correspond to somewhat higher concurrence values.

In the alternative scenario, the virtual source PDL $\vec{\gamma}^S$ is placed in channel B while the emulator PDL $\vec{\gamma}^E$ is in channel A, as depicted in Fig. 2(b). Now channel B has a constant PDL, while the PDL of channel A is variable in magnitude and orientation. The latter can be characterized by $\kappa = (T\hat{\gamma}^S) \cdot \hat{\gamma}^E$. It follows from Eq. 2 that the correlation matrix element $t_3 = 1$, because our input state is $|\Phi^+\rangle$. Interestingly, since the vector $\vec{\gamma}^S$ points in the $\sigma_3$ direction, $\kappa$ can be easily related to the measured angle $\vartheta$ described above as: $\kappa = (T\hat{\gamma}^S) \cdot \hat{\gamma}^E = \hat{\gamma}^S \cdot \hat{\gamma}^E = \cos\vartheta$. Fig 2(d) shows the same experimental data seen in Fig. 2(c), but now the concurrence is shown as a function of both the magnitude of applied PDL ($\gamma^E$) and the relative orientation of two vectors $\vec{\gamma}^S$ and $\vec{\gamma}^E$. The surface is calculated based on Eq. 3. Again, the data clearly supports our theoretical results. For $\kappa = -1$, the PDL of both channels are counter aligned such that they partially cancel each other, and the concurrence is maximized (for $\gamma^E = \gamma^S$). On the other hand, $\kappa = +1$ corresponds to the case where the PDL of each channel is aligned in the same direction, thus increasing the effective PDL and minimizing concurrence.

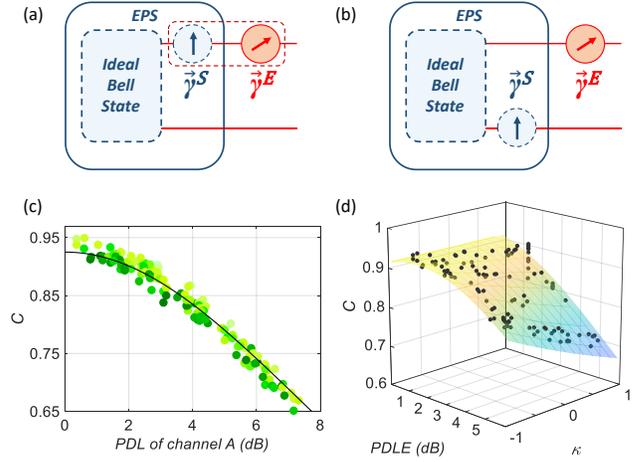

**Fig. 2** (a, b) The EPS of Fig. 1 is modeled by a combination of an ideal Bell State source and a virtual PDL $\vec{\gamma}^S$, which can be positioned either in channel A (a) or channel B (b). $\vec{\gamma}^E$ denotes the PDL vector produced by the emulator PDLE in channel A. (c) Shaded circles: concurrence C vs PDL of channel A (both computed from the measured tomography). Color denotes purity with lighter shades corresponding to higher purity values. Black line shows the theoretical dependence of Eq. 5. (d) Black circles: the same values of concurrence C as in (c) but presented as a function of the magnitude of PDL of the emulator PDLE and the value of $\kappa$ (Eq. 4). The values of C and $\kappa$ are computed from the measured tomography. The surface is calculated based on Eq. 3.

## 4. NONLOCAL POLARIZATION DEPENDENT LOSS COMPENSATION

An arbitrarily oriented PDL element $\vec{\gamma}_A$ in channel A can be mapped to a certain PDL element $\vec{\gamma}_B$ in channel B. In other words, for a two photon state resulting from application of $\vec{\gamma}_A$ on a Bell state, an equivalent two photon state can be obtained by applying a properly chosen $\vec{\gamma}_B$ to the same initial Bell state. That is, the two density matrices post-selected by the coincidence measurement are equal. A general relationship for finding *the equivalence mapping* of an arbitrarily oriented PDL element from one qubit to the other can be found by solving the equation

$$(P_1 \otimes \sigma_0)\rho(P_1 \otimes \sigma_0)^\dagger = (\sigma_0 \otimes P_2)\rho(\sigma_0 \otimes P_2)^\dagger. \quad (6)$$

We find that when the initial state is a Bell state, identical density matrices result from replacing PDL on one qubit of magnitude $\gamma_A$ and orientation $\hat{\gamma}_A$ in Stokes space with an element of equal magnitude and different orientation given by $\hat{\gamma}_B = T\hat{\gamma}_A$ in channel B (shown schematically in Fig. 3(b)). Here, $T$ is the correlation matrix. Since all elements of $T$ satisfy $|t_j| = 1$ for Bell states, application of $T$ to a Stokes vector constitutes an inversion of some or all axes in Stokes space. Fig. 3(a) illustrates these transformations for an arbitrary unit vector $\hat{\gamma}_A$, which is shown on the sphere (red) together with four equivalently mapped vectors $\hat{\gamma}_B$ (green). Each vector $\hat{\gamma}_B = T\hat{\gamma}_A$ corresponds to a particular input Bell state and is labeled accordingly. When the input state is a singlet, the inversion occurs in all three axes; however, only one axis is inverted for each of the triplets.

The existence of an *equivalence mapping* suggests that PDL can be compensated nonlocally. Locally, two concatenated PDL elements of equal magnitude and anti-parallel Stokes vectors in *the same channel* simply reduce to pure loss which attenuates each mode equally. In a more general case, one of these elements could be considered as an image mapped from a real element in the other channel. Then, to compensate for a PDL element of magnitude $\gamma_A$ and orientation $\hat{\gamma}_A$ in channel A, we use an additional element in channel B with magnitude $\gamma_B = \gamma_A$ and

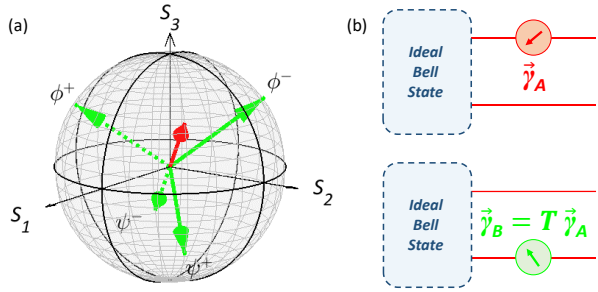

**Fig. 3** (a) An arbitrary real unit vector in channel A (red) together with four virtual vectors in channel B (green), each corresponding to a particular input Bell state, given by its label. (b) *Equivalence mapping* of a real PDL element $\vec{\gamma}_A$ in channel A onto a virtual PDL element $\vec{\gamma}_B = T\vec{\gamma}_A$ in channel B for an ideal Bell state source.

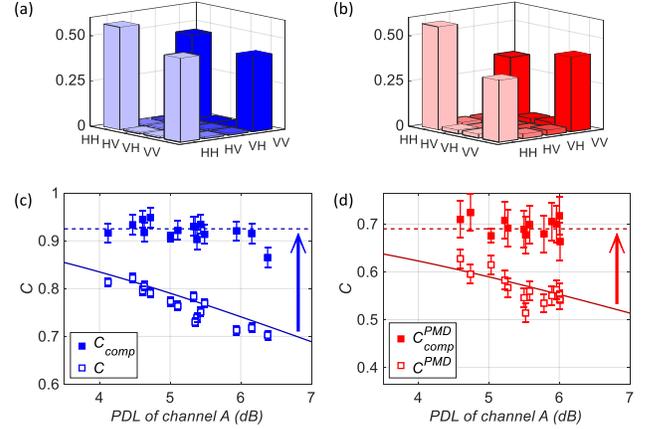

**Fig. 4** (a) Absolute value of a back-to-back density matrix with PDLE/PDLC removed. (b) Absolute value of a density matrix with PMD ($\tau \sim 6.6$ ps) in channel A and no applied PDL. (c) Concurrence C vs PDL of channel A (both computed from the measured tomography). Open symbols denote the uncompensated case with PDLC removed, and closed symbols mark the maximum achieved concurrence. (d) Data similar to that shown in (c) but with additional PMD-induced decoherence in channel A as in (b).

orientation $\hat{\gamma}_B$ such that $T\hat{\gamma}_B \cdot \hat{\gamma}_A = -1$. The latter indicates that in channel A, the real element $\hat{\gamma}_A$ and the element mapped from channel B ($T\hat{\gamma}_B$) are anti-parallel. Indeed, it follows from Eq. 3 that when these conditions are satisfied, $C(\rho') = C(\rho)$ and complete PDL compensation is achieved nonlocally.

Now we modify our setup to demonstrate PDL compensation. We keep the same PDL module in channel A, called the "emulator" (red PDLE box in Fig. 1(a)), and add a functionally similar PDL module in channel B as the "compensator" (green PDLC box in Fig. 1(a)). The following procedure is employed to vary the "PDL of channel A." With the magnitude $\gamma^E$ of the emulator PDL $\vec{\gamma}^E$ being set to a fixed value of 5.1 dB, we change its orientation $\hat{\gamma}^E$ using the polarization controller PCE. As the "PDL of channel A" results from the concatenation of the virtual source PDL $\vec{\gamma}^S$ and the PDL of the emulator $\vec{\gamma}^E$ (Fig. 2(a)), we obtain a range of the magnitude of the "PDL of channel A" from 4.1dB to 6.4dB. Notice that this range indicates that we cover nearly all possible angles $\vartheta$ between $\hat{\gamma}^E$ and $\hat{\gamma}^S$ within our experimental accuracy. For each of the values of PDL in channel A, we set the emulator in channel B to zero, perform tomography, and calculate the concurrence, which is plotted as empty symbols in Fig. 4(c). These uncompensated results are similar to those of Fig. 2(c). In fact, the solid lines in both figures plot exactly the same dependence of Eq. 5. Then we set the magnitude of the PDL of the compensator PDLC to be equal to the experimentally established value of the PDL of channel A: $\gamma_B = \gamma_A$. By searching through various orientations of PDLC $\hat{\gamma}_B$, we find the one resulting in maximal concurrence, which is plotted with filled symbols in Fig. 4(c). The horizontal dashed line marks a level of $C = 0.925$, corresponding to the measured B2B concurrence without any PDL in either channel. As the plot shows, we are able to recover the initial concurrence for all experimentally available values and orientations of PDL in channel A within our experimental accuracy.

As our final setup modification, we keep the compensating PDL module in channel B intact (green PDLC box in Fig. 1(a)) while replacing the emulator PDL module in channel A (red PDLE box in Fig. 1(a)) with a different unit that has a fixed amount of first order PMD (differential group delay of $\tau = 6.6$ ps) in addition to the variable PDL. First order PMD partially reduces the input state coherence [23-24], thus transforming a nearly perfect $|\Phi^+\rangle$ Bell state into a rank two Bell diagonal state $\rho = \frac{1+C}{2}|\Phi^+\rangle\langle\Phi^+| + \frac{1-C}{2}|\Phi^-\rangle\langle\Phi^-|$, where $C$ is the resulting Bell diagonal state concurrence. To characterize this state experimentally, we perform quantum state tomography with the PDL magnitudes of both the emulator and compensator dialed to zero: $\gamma_A = \gamma_B = 0$. Fig. 4(b) presents the resulting density matrix, while Fig. 4(a) shows a typical density matrix obtained from the B2B measurements without any PMD present. The reduced coherence due to PMD exhibits itself in the smaller off-diagonal elements of the matrix in the right panel and the correspondingly smaller value of computed concurrence given by $C = 0.69$. As described earlier, a slight imbalance between the $|HH\rangle$ and $|VV\rangle$ modes arises from the intrinsic source PDL $\vec{\gamma}^S$.

To investigate nonlocal PDL compensation in the presence of decoherence, we followed the same protocol as described above. The PDL of channel A was varied by adjusting the Stokes angle $\vartheta$ between the emulator PDL $\hat{\gamma}^E$ and the intrinsic source PDL $\hat{\gamma}^S$. For each case, we first took uncompensated tomography data by keeping the compensator value at zero, $\gamma_B = 0$. Then $\vec{\gamma}_B$ was adjusted both in magnitude and direction to achieve the best PDL compensation possible. The uncompensated (compensated) concurrence values are shown by empty (filled) symbols in Fig. 4(d). The horizontal dashed line denotes $C = 0.69$, which corresponds to the concurrence of the PMD decohered state without any PDL applied to either channel (shown in Fig. 4(b)). Our experiment proves that the initial concurrence is again largely restored for all experimentally available magnitudes and orientations of PDL in channel A. Note that for a general mixed state, complete PDL compensation is not necessarily possible. We speculate that alignment of the PDL and PMD vectors in our emulator might enable the complete compensation that we have observed.

## 5. AVERAGE ENTANGLEMENT AND TRANSMISSION RATES

Intuitively, the introduction of an extra loss element $\vec{\gamma}_B$ could only reduce the transmission rate of entangled pairs, even when the combined action of such an element with $\vec{\gamma}_A$ leads to a higher degree of entanglement of the pairs. Let us consider the average entanglement [28] as the product of the concurrence $C(\rho')$ of the final density matrix and the transmission rate of detected pairs: $\Gamma = \text{Tr}[(P_A \otimes P_B)\rho(P_A \otimes P_B)^\dagger]$. Again adopting the concurrence expression of Ref. [29] to PDL operators $P_A$ and $P_B$ in the form of Eq. 1, we find that the average entanglement is given by:

$$\Gamma C(\rho') = C(\rho)|\det(P_A)||\det(P_B)| = e^{-(\gamma_A+\gamma_B)}C(\rho), \quad (7)$$

where $C(\rho)$ is the concurrence of the initial state. It follows from Eq. (7) that the average entanglement depends only on the total sum of PDL magnitudes in each channel. Neither individual PDL orientation nor its partition between the channels affects the average entanglement. In other words, given a two qubit system with PDL of *fixed magnitude* in each channel, there is a trade-off between entanglement quality, such as measured by concurrence, and the transmission rate of entangled pairs. This trade-off is controlled by the relative orientation of $\hat{\gamma}_A$ and $T\hat{\gamma}_B$.

The normalized concurrence and corresponding normalized pair transmission rates are shown in Fig. 5(c, d) as a function of the PDL of channel A. The experimental dataset consists of 12 different settings of PDL of channel A out of about 30 shown in Fig. 4. For each of the settings, the PDL of channel B is set to $\gamma_B = \gamma_A$, while the direction of $\hat{\gamma}_B$ is varied through approximately 15 different orientations. The data for six of these settings of the PDL of channel A is taken with PMD-induced decoherence. The concurrence measured at *each* $\hat{\gamma}_B$ orientation together with the corresponding rates are plotted in Fig. 5 with filled circles colored by shades of blue according to the concurrence values. Here, both the concurrence and the rate are normalized by their values with no PDL applied. In contrast, filled symbols in Fig. 4 (c, d) plot only the maximum attained values that correspond to the optimum orientation of $\hat{\gamma}_B$. The color shading indicates the orientation of $\hat{\gamma}_B$, which ranges from $\kappa = T\hat{\gamma}_B \cdot \hat{\gamma}_A = -1$ (lighter shades) to $\kappa = T\hat{\gamma}_B \cdot \hat{\gamma}_A = 1$ (darker shades). That is, the lighter shades denote the PDL compensation scenario, and hence exhibit the maximum concurrence and lowest rates. Corresponding theoretical expressions are plotted with the lightly colored lines: $C_{\kappa=-1} = 1$ and $\Gamma_{\kappa=-1} = \frac{1}{2}(e^{-2\gamma_A} + e^{-2\gamma_B})$. The darker shades mark the opposite situation of minimum concurrence and maximum transmission rate. The dark lines on the plots are theoretical curves for the minimum concurrence $C_{\kappa=+1} = \text{sech}(\gamma_A + \gamma_B)$ and the maximum rate $\Gamma_{\kappa=+1} = \frac{1}{2}(1 + e^{-2(\gamma_A + \gamma_B)})$. The concurrence data fit the theory rather well, while the rate data exhibits larger errors mostly due to the unfiltered Raman noise of our source. Yet the tradeoff between concurrence and rate is quite obvious for all magnitudes and orientations with and without PMD degradation.

Finally, as an aside we examine reduced density matrices. Consider a dataset taken at a specific value of 5.27 dB of PDL in channel A (marked by the arrow in Fig. 5(c)). Panel (a) plots the concurrence of this slightly decohered two qubit state (of Fig. 4(b)) vs linear entropy of its qubit A for this dataset. In addition, panel (b) shows three reduced density matrices for qubit A that correspond to the entropy values marked by squares in Fig. 5(a). Indeed, it is clear from Fig. 5(a) that as the entropy of photon A is increased and the reduced density matrices gradually become depolarized, the orientation of $\hat{\gamma}_B$ approaches its optimum value for compensation, and the concurrence increases. It was predicted theoretically that a two photon mixed state reaches it maximum entanglement under local filtering operations when the reduced density matrices approach the identity matrix [27, 47, 48]. Our data is an explicit experimental demonstration of that theory. It further suggests that this depolarization of an individual qubit could serve as the feedback for practical PDL compensation.

## 6. CONCLUSION

Here we have considered practical fiber-optic quantum channels that partially filter individual modes of transmitted polarization entangled states by channels' polarization dependent loss. For these channels, we have introduced a compact model that characterizes how the concurrence of Bell diagonal states is affected by arbitrarily oriented PDL. Then, by using a fiber-based polarization entanglement distribution system which operates at telecom wavelengths, we were able to demonstrate several effects predicted by this model. These

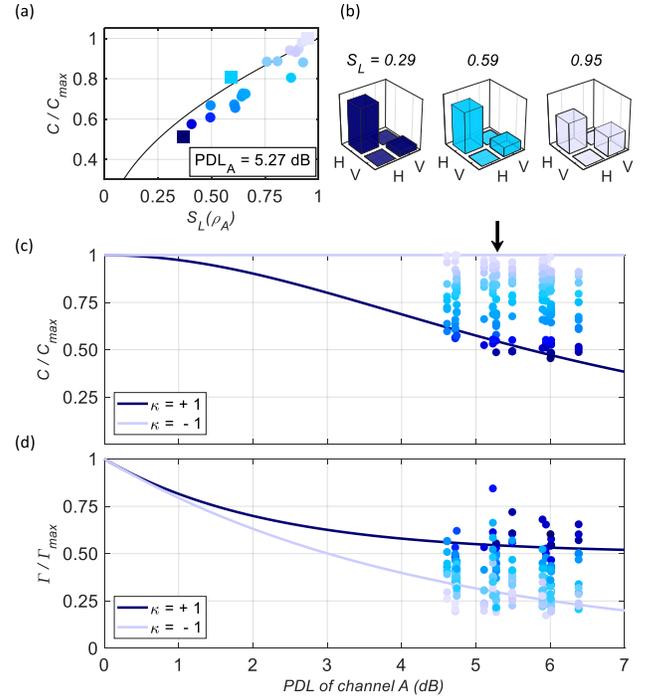

**Fig. 5** (a) Concurrence of a partially decohered two photon state for different orientations of $\hat{\gamma}_B$, plotted as a function of the linear entropy of photon A. PDL of channel A is equal to 5.27 dB and PMD in channel A is $\tau \sim 6.6$ ps. The black arrow points to the same concurrence data in panel (c). Three squares mark points with linear entropy values of $S_L = 0.29, 0.59,$ and $0.95$. (b) Density matrices computed for the points marked by squares. (c) The normalized concurrence measured at different orientations of $\hat{\gamma}_B$ plotted as a function of the PDL of channel A. Color encodes concurrence values throughout this figure. The lines plot minimum and maximum theoretical concurrence. (d) The normalized transmission rate of entangled pairs shown for all data portrayed in (c). Points of a given color in (d) correspond to points of the same color in (c). The lines plot minimum and maximum theoretical rates.

include the dependence of concurrence on PDL orientation and the nonlocal compensation of PDL with and without introducing additional PMD-induced decoherence. We also described the conservation of the average entanglement during nonlocal compensation. The resulting tradeoff between concurrence and transmission rate provides insights into the capacity of practical fiber-optic channels. Thus, our results pave the way toward the fiber-optic realization of quantum networks.

**Acknowledgment**. We would like to acknowledge stimulating discussion with W. J. Munro.